\documentclass[preprint,amsfonts,showpacs,nofootinbib]{revtex4}
\usepackage{mathptmx}
\usepackage[latin1]{inputenc}
\usepackage{amsfonts}
\usepackage{amssymb}
\newcommand{\be}{\begin{equation}}
\newcommand{\ee}{\end{equation}}
\def\sot#1#2{\mathop{\vtop{\ialign{##\crcr
$\hfil\displaystyle{#2}\hfil$\crcr\noalign
     {\kern1pt\nointerlineskip}\hbox{$\hfil{{}_{{}_{#1}}}\hfil$}\crcr
     \noalign{\kern1pt}}}}}
\newcommand{\sv}{{v_{\rm S}}}
\newcommand{\ps}{{p_{\rm S}}}
\newcommand{\tv}{{v_{\rm T}}}
\newcommand{\pt}{{p_{\rm T}}}

\newcommand{\ba}{\begin{eqnarray}}
\newcommand{\ea}{\end{eqnarray}}

\newtheorem{prop}{Proposition}[section]
\newtheorem{defin}{Definition}[section]

\begin{document}

\title{A generalized Principle of Relativity}
 \author{Fernando de Felice}
 \email{fernando.defelice@pd.infn.it}
\author{Giovanni Preti}
 \email{giovanni.preti@pd.infn.it}
\affiliation{Dipartimento di Fisica ``Galileo Galilei", Universit\`a degli Studi di Padova, and INFN Sezione di Padova --- Italy}

\begin{abstract}
The Theory of Relativity stands as a firm groundstone on which modern physics is founded.
In this paper we bring to light an hitherto undisclosed richness of this Theory, namely its admitting a consistent reformulation which is able to provide a unified scenario for all kinds of particles, be they lightlike or not.
This result hinges on a {\em generalized} Principle of Relativity which is {intrinsic} to Einstein's Theory --- a fact which went completely unnoticed before. 
The road leading to this generalization starts, in the very spirit of Relativity, from enhancing full equivalence between the four spacetime directions by requiring full equivalence between the {\em motions} along these four spacetime directions as well.
So far, no measurable spatial velocity in the direction of the time axis has ever been defined, on the same footing of the {\it usual} velocities -- the ``space-velocities" --- in the local three-space of a given observer. 
In this paper, we show how Relativity allows such a ``time-velocity" to be defined in a very natural way, for any particle and in any reference frame. 
As a consequence of this natural definition, it also follows that the time- and space-velocity vectors sum up to define a spacelike ``world-velocity" vector, the modulus of which --- the {\it world-velocity} --- turns out to be equal to the Maxwell's constant $c$, irrespective of the observer who measures it. 
This {measurable} world-velocity (not to be confused with the space-velocities we are used to deal with) therefore represents the speed at which {\em all} kinds of particles move in spacetime, according to {\em any} observer.
As remarked above, the unifying scenario thus emerging is intrinsic to Einstein's Theory; it extends the role traditionally assigned to Maxwell's constant $c$, and can therefore justly be referred to as ``a generalized Principle of Relativity".

\end{abstract}


\maketitle

\section{Introduction}
``Velocity" is a {\em relative} concept; be it the velocity of light or the velocity of a nonluminal
particle, there always exists an {\em observer} who {\em measures} it with a suitable device
and in a given frame.
Yet, the Principle of Relativity is grounded on the well established experimental fact that the vacuum light velocity $c$
 has an
{\em absolute} character, being independent of the observer.
This crucial experimental result --- wholly unexpected, at Michelson and Morley's times --- forced the well known drastic change
of perspectives: new transformations had to be worked out, superseding the Galileian ones, in order to cope
with the experimental findings, and the informations codified in these new ``Lorentz" transformations opened
the scenario for the physical revolution which ultimately led to Einstein's Relativity.
Existence of the absolute constant $c$ within the theory of Relativity allows {unambiguous} conversion
of time units into length ones; resting on this fact and on the experimentally
 well-tested Lorentz transformations, a simple and direct procedure will lead us to discover that
the Maxwell constant $c$ plays a broader physical role than usually credited to it.
It represents the universal, observer independent ``world-velocity", characterizing {\em any} physical
entity according to {\em any} physical observer.

For the sake of argumentation, the subject will initially be approached from a Special Relativistic
point of view and employing the simplest observers; yet, the intrinsically covariant character
of the formulae which will be derived guarantees that the results hold in any spacetime, curved or not.
Hence, the universal character of the world-velocity is an intrinsic --- yet hitherto unnoticed --- content of Einstein's theory.

In the theory of Relativity, time $t$ has the status of a {coordinate} which spans,
together with the three spatial ones, a four-dimensional manifold endowed with a pseudo-Riemannian geometry,
termed ``spacetime".
Due to basic dimensional uniformity requirements,
the time coordinate must have the dimension of a length --- which requires $t$ to be converted into $ct$.
Such a conversion is {\em unambiguous}, since experiment definitely certifies the universal character
of the conversion factor $c$ \cite{TW} \cite{W}.
As a consequence, time intervals can unambiguously be converted into spatial ones,
and therefore a ``measurable spatial velocity along the time axis" can most naturally be defined,
on the same footing of the ``usual" velocities along the three space axes.
Introduction of this ``time-velocity", as we shall call it, is not pure aesthetics:
once the existence of this {\em physically measurable} velocity is recognized,
it follows as a natural outcome that a {\em spacelike} four-velocity exists,
the physically measurable modulus of which --- the above mentioned
world-velocity --- is identically equal to Maxwell's constant $c$.
This evidences an underlying unity in the behaviour of {\em all} kinds of particles.
The universality of the world-velocity provides a rule to which all physical entities conform;
it extends Einstein's Principle of Relativity regarding the universality of $c$ as the velocity of lightlike particles, and can therefore be regarded as a generalized Principle of Relativity. 
Such a generalization, being intrinsic to Einstein's Theory, does most obviously {\em not} represent a proposal for a new theory, nor the seed for an alternative structure: strongly advocating Einstein's Relativity, in this paper we are bringing to light a richness of its which had never been explored before. 

The contents of the paper are organized as follows:
 in section 2 the concepts of space-, time-, and world-velocities are introduced, and the generalized Principle is enunciated;
space and time comotions are examined in section 3;
section 4 provides a discussion about the role of mass in Relativity, based on the relation existing between mass and time-velocity; section 5 is dedicated to the issues of space- and
time-momenta and energy of a generic real particle; section 6 is devoted to an application of the concept of T-velocity to the twin problem, making its solution almost straightforward.
The concluding section 7 will finally recall the main points of the paper.

Notation: Greek (``spacetime") indices run from 0 to 3, Latin (``space") ones run from 1 to 3.
 The $c$ factors are explicitly included in the formulae throughout.

\section{Space-, time-, and world-velocities}
In order to make things as simple as possible,
we begin our analysis in the flat spacetime environment of Special Relativity spanned by the metric
\be \label{metric}
ds^2=-c^2t^2+dx^2+dy^2+dz^2,
\ee
and consider the
simplest couple of non comoving observers, namely the static observer $\bf u$, identified by the
four-velocity
\be
\label{obss}
u^\alpha=c\delta^\alpha_0
\ee
and the inertial observer $\bf u'$, moving with respect to $\bf{u}$ at the {\em constant} instantaneous  space velocity
$c\beta=c\sqrt{\delta_{jk}{\beta^j\beta^k}}$ and  Lorentz factor $\gamma=1/\sqrt{(1-\beta^2)}$, viz.
\be
\label{obs}
u'{}^{\alpha}=c\gamma(\delta^\alpha_0+\beta^j\delta^\alpha_j).
\ee
Both the four-velocities (\ref{obss}) and (\ref{obs}) describe {\em timelike} motions; hence,
they are normalized according to the rule
\be
\label{norm}
u_\alpha u^\alpha=u'_\alpha u'{}^{\alpha}=-c^2,
\ee
which actually {\em defines} the four-velocity as the tangent vector along the worldline when
proper time is selected for its parametrization.

If we chose the observer $\bf u$ as the one who makes the measurements then we identify the spacetime with metric (\ref{metric}) as ``the {\em frame} of $\bf u$", and ``the {\em observer $\bf u$}" as the family of such static observers each being fixed at a spatial position and having proper time running with the same rate; hence, at each point of the frame of $\bf u$ there exists a $\bf u$-clock running with the
time $t_u$.
The observer $\bf u'$ instead, being observed by $\bf u$, moves in the frame of $\bf u$ along a trajectory --- described by (\ref{obs}) --- which crosses a $\bf u$-clock at each of its points. The above considerations are essential to the definitions of velocities ``{\em with respect to $\bf u$}" we shall introduce shortly.

The covariant expression for the {\em three}-dimensional velocity $c\beta^j$ appearing
in eq.(\ref{obs}) is given by the following {\em four}-vector:
\be
\label{svalpha}
\sv(u',u)^\alpha\equiv-\frac{c^2P(u)^\alpha{}_\beta u'{}^{\beta}}{(u_\sigma u'{}^{\sigma})}=
\frac {u'^{\alpha}}{\gamma} - u^\alpha,
\ee
obtained by projecting $\bf u'$ on the local rest space of $\bf u$ with the projection operator \cite{dFC}
\cite{JCB}
\be
\label{Pop}
P(u)^\alpha{}_\beta=\delta^\alpha_\beta+u^\alpha u_\beta/c^2.
\ee
The Lorentz factor $\gamma$ has its own covariant form as well, given by
\be
\label{gamma}
\gamma=-(u_\alpha u'{}^{\alpha})/c^2.
\ee
Due to the spatial character of the projector (\ref{Pop}), the vector $\sv(u',u)^\alpha$
defined in eq.(\ref{svalpha}) is orthogonal to ${\bf u}$,
which means that it lies in the local three-space of this observer.
Physically, $\sv(u',u)^\alpha$ is the space-velocity vector (clearly, a spacelike quantity) of ${\bf u'}$
as determined by the measurements made by ${\bf u}$.
Note, at this regard, that both the space and the time intervals characterizing
this velocity are correctly expressed in terms of the space
and time of $\bf u$; in coordinates: $\sv^j=d{x}^j/dt\,=c\beta^j$.

The coordinate-independent quantity associated with the space motion of $\bf u'$ which the
observer $\bf u$ can determine via local measurements is the {\em modulus} $\sv(u',u)$ of the
four-vector (\ref{svalpha}); taking into account (\ref{gamma}),
this modulus reads
\be\label{sv}
\sv(u',u)=\sqrt{\sv(u',u)_\alpha\, \sv(u',u)^\alpha}=c\sqrt{1-\frac{1}{\gamma^2}}.
\ee
We can now give the following
\begin{defin}
\label{def1}
The quantity ${\sv}(u',u)$ introduced in eq.(\ref{sv}) is termed  the
{\em space-ve\-loc\-ity} --- or {\em S-velocity}, for short --- of ${\bf u}'$
with respect to $\bf u$.
\end{defin}
As the above definition implies, the S-velocity ranges from $0$ to $c$
(asymptotically, for ultrarelativistic particles);
in the limit case, the well known lightlike behaviour $\sv=c$ is recovered.

While moving along the three local space axes with velocity $\sv^j$ along each of them,
$\bf u'$ is also observed by $\bf u$ to move along the temporal axis, as eq.(\ref{obs}) clearly shows;
indeed, this motion is in general much more rapid, unless $\bf u'$  moves at relativistic
speed (i.e., unless $\sv\rightarrow c$).
To a time interval along the local time axis (i.e., the one-dimensional manifold parametrized  by the
proper time along each
observer's worldline) there unambiguously corresponds, as remarked in the introduction,
a length interval along the time coordinate
(which is dimensionally a length, and therefore identifies a {\it space} in the
conventional sense of this term).

As the observer $\bf u'$ moves in the spacetime of $\bf u$, he will read a time $t_{u}$ on each $\bf u$-clock  he crosses during his motion. He will then compare that time with the time $t'_{u'}$ read on his own clock at the same instant.
 For any two infinitely close points
 on the trajectory of $\bf u'$,  we have the relation
\be\label{t't}
dt'_{u'}=\sqrt{1-\beta^2}d t_{u}.
\ee
While $dt'_{u'}$ is an interval of the proper time of the observer $\bf u'$, the quantity $dt_{u}$
is just the difference of the time read by $\bf u'$ on two different $\bf u$-clocks which are being crossed by $\bf u'$ during his motion, as stated.
Due to the existence of the {\em single and universal} time--length conversion
factor represented by the Maxwell constant $c$, the above relation implies that in the time
interval $d t_{u}$ read on two subsequent $\bf u$-clocks, the observer $\bf u'$ has
covered a {\em spatial} distance
$cd t'_{u'}$, as specified in (\ref{t't}), along the {\em time dimension} of $\bf u$.
To this distance there naturally corresponds an instantaneous velocity
\be
\label{tv}
\tv(u',u)=\frac{cd t'_{u'}}{d t_{u}}=\frac c\gamma=
-\frac{c^3}{u_\alpha u'{}^{\alpha}}
\ee
assigned by $\bf u$ --- as the result of physical measurements he has made --- to the time motion of $\bf u'$.
Existence of this velocity is a plain consequence of the existence of the Maxwell constant $c$ and of the Lorentz transformations.
Equation (\ref{tv}) defines a covariant scalar quantity, at which regard we can give the following
\begin {defin}
\label{def2}
The quantity $\tv(u',u)$ introduced in eq.(\ref{tv}) is termed the
 {\em time-ve\-loc\-i\-ty} --- or {\em T-ve\-loc\-i\-ty}, for short ---
of ${\bf u'}$ with respect to $\bf u$.
\end{defin}
We remark that the T-velocity, similarly to the S-velocity,
is a {\em physically measurable} quantity; its value can be
obtained from a local measurement of the covariant $\gamma$ factor (\ref{gamma}).
 From its defining equation (\ref{tv}), it is clear that the T-velocity ranges
from $\tv=c$, when $\gamma=1$, to $\tv\to 0$, when $\gamma\to\infty$;
a ultrarelativistic particle
{asymptotically} approaching a lightlike behaviour is therefore {observed} to approach a
vanishingly small T-velocity.

Introduction of the concept of a ``time-velocity" allows a natural conversion of the traditional idea
of ``time evolution with respect to a given observer" (the ``relative aging", broadly speaking)
into the idea of a {\em spatial} motion along the time direction, on the same
footing of the ``ordinary" motions in the local three-{\em space} of a given observer.
Thus, perfect equivalence among the spacetime coordinates is restored, in the spirit of Relativity.

Alike the S-velocity, the T-velocity too can be expressed in a natural way as the modulus of
a four-vector, via a simple projection procedure.

\noindent First, we recall that the time interval $dt_u$, measured by the observer $\bf u$ and
 corresponding to a displacement
$d{x}^\alpha=u'{}^{\alpha}dt'_{u'}$  along the world-line of $\bf u'$, is given by \cite{dFC} \cite{JCB}
\be
\label{dtu}
dt_u=-\frac{1}{c^2}u_\alpha dx^\alpha=-\frac{1}{c^2}u_\alpha u'{}^{\alpha}dt'_{u'}
\ee
(evidently, this relation is identical to (\ref{t't}), since the observer who makes
the measurement is still $\bf u$).

\noindent Second, using the projection operator on the
time axis of $\bf u$, namely
\be
\label{Pip}
\Pi(u)^\alpha{}_\beta=-u^\alpha u_\beta/c^2,
\ee
we can rewrite
 eq.(\ref{dtu}) in the following form:
\be
\label{dt2u}
dt^2_u=-\frac1{c^2}\Pi(u)^\alpha{}_\beta u'_{\alpha}u'^{\beta}d{t'}^2_{u'}.
\ee
Third, recalling that $d{t'}_{u'}=d{t}_{u}/\gamma$, and the properties of projectors, eq.(\ref{dt2u})
can be rewritten as
\be\label{pp}
dt^2_u=-\left(\Pi(u)^{\alpha}{}_{\sigma}u'{}_{\alpha}\frac{dt_u}{c\gamma}\right)\left(
\Pi(u)^\sigma{}_{\beta}u'{}^{\beta}\frac{dt_u}{c\gamma}\right).
\ee
Fourth, recalling that $\tv=c/\gamma$, we easily find from (\ref{pp})
$$
\tv^2=\frac{ic^4\Pi(u)^{\alpha}{}_{\sigma}u'{}_{\alpha}}{(-u_\rho u'{}^\rho)^2}\cdot\frac{ic^4\Pi(u)^\sigma{}_{\beta}u'{}^{\beta}}{(-u_\rho u'{}^\rho)^2}\equiv\tv_\sigma\tv^\sigma,
$$
where
\be
\label{tvalpha}
\tv(u',u)^\alpha\equiv-\frac{ic^4\Pi(u)^\alpha{}_\beta u'{}^\beta}{(-u_\rho u'{}^\rho)^2}=\frac{i\,u^\alpha}{\gamma}
\ee
is the sought-for T-velocity vector, which is {\em spacelike} and orthogonal to the spacelike S-velocity vector $ \sv(u',u)^\alpha$ of eq.(\ref{svalpha}) due to the intrinsic properties of the projection operators (\ref{Pop}) and (\ref{Pip}).
The T-velocity four-vector (\ref{tvalpha}), being {spacelike}, is naturally adequate to describe a {spatial} velocity;
note that its imaginary character rises no problems, since (\ref{tvalpha}) is {\em not} a measurable quantity, at variance with the T-velocity (\ref{tv}), which is {\em the} physical observable.

The two {spacelike} velocity vectors $\sv(u',u)^\alpha$ and $\tv(u',u)^\alpha$, obtained via simple projection procedures, can now be vectorially summed:
\be
\label{wv}
w(u',u)^\alpha\equiv  \sv(u',u)^\alpha + {}\tv(u',u)^\alpha
\ee
into a new {\em spacelike} velocity vector, the modulus of which we now wish to calculate.
At this regard, note that definition (\ref{wv}) does {\em not} introduce a complexification of the tangent space, since $\sv(u',u)^\alpha\in\mathbb{R}^3$, $\tv(u',u)^\alpha\in i\mathbb{R}$
and $w(u',u)^\alpha\in\mathbb{R}^3\oplus i\mathbb{R}$, isomorphic to $\mathbb{R}^4$ and not to $\mathbb{C}^4$ (which is isomorphic to $\mathbb{R}^8$).
Hence, we do not have to introduce a complexification of the metric real tensor into a complex Hermitian one in order to preserve the necessary symmetry of the scalar product:
the metric remains unchanged --- the ``old" real one --- and the modulus $w(u',u)$ of (\ref{wv}) is correctly calculated with the usual rule, which gives:
\be
\label{totalv}
w(u',u)=\sqrt{w(u',u)_\alpha w(u',u)^\alpha}=\sqrt{\sv(u',u)^2+\tv(u',u)^2} \equiv c
\ee
{\em identically}, irrespective of the choice of the
couple $\{\bf{u},\bf{u'}\}$.
Thus, once we introduce the following
\begin{defin}
\label{def3}
The quantity $w(u',u)$ defined in eq.(\ref{totalv}) is termed
the {\em world-velocity} of $\bf u'$ with respect to $\bf u$.
\end{defin}
we can state the
\begin{prop}
\label{pr1}
The world-velocity of any {\em nonluminal}  particle
is equal to the Maxwell constant $c$ with respect to {\em any} observer who measures it.
\end{prop}
We recall that both the space- and the time-velocities $\sv$ and $\tv$ are {\em physically measurable}
quantities; so therefore is their composition (\ref{totalv}) into the world-velocity.
We have just seen that the value of the world-velocity characterizing the nonluminal particles turns out to coincide with a universal constant, Maxwell's $c$, which is normally associated with lightlike motions alone.
This result suggests the existence of a unifying rule holding for {\em all} real particles, be they lightlike or not, and leads naturally to the enunciation of a generalized Principle of Relativity, as we are going to see.

Before proceeding further, yet, a remark is due.
The physical content of the relation $w_\alpha w^\alpha=c^2$, cf.\,eq.(\ref{totalv}), for the spacelike {world}-velocity
vector $\bf w$ is {\em not} a simple restatement of the information already encoded in the normalization condition $u_\alpha u^\alpha=-c^2$, eq.(\ref{norm}),
holding identically for {any}
{\em timelike} velocity four-vector $\bf u$.
In fact, the normalization condition represents just the {\em definition} of the four-velocity (see above), but {\em it does not involve any sort of measurement}; in particular, it does {\em not} imply that the given $\bf u$ is spatially moving
at velocity $c$ --- not any more, for sure, than the corresponding lightlike particle relation $k_\alpha k^\alpha=0$ implies that a light signal has zero spatial momentum --- relative to any observer.

Proposition \ref{pr1} deals with nonluminal particles; even if no observer can be associated with lightlike trajectories, we have already noted that the limit case of eq.(\ref{sv}) as $\gamma\rightarrow\infty$ agrees with the experimentally established observer-independent $\sv=c$ value.
The same limit procedure, when applied to eq.(\ref{tv}), shows that lightlike particles should be assigned the T-velocity $\tv=0$, identically; correspondingly, eq.(\ref{totalv}) would be satisfied in this limit case too.
Consistency of this limit procedure justifies the ansatz that a more general version of proposition \ref{pr1}
can be given, namely
\begin{prop}[{\it or:} ``The generalized Principle of Relativity'' ]
\label{pr2}
The world-velocity of {\em any} particle
is equal to the Maxwell constant $c$ with respect to {\em any} observer who measures it.
\end{prop}
Because of its generality and the novelty of its content, proposition \ref{pr2} can be conceived as a generalized Principle of Relativity.
It attributes to {\em all} particles the property of moving in spacetime with an observer-independent {measurable} world-velocity equal to $c$.
The more familiar lightlike particle property of moving in spacetime with an observer-independent measurable space-velocity equal to $c$ (the ``old'' Principle) is thus seen to represent just a particular case of this generalized Principle.

Some implications of this generalized Principle are discussed in the next sections.

\section{Space and time comotion}

The relative character of a velocity naturally implies the concept of ``comotion'', realized when the relative velocity between two observers vanishes.
While this concept is quite obvious in the case of a space-velocity, it is not so in the case of a time-velocity.
In order to clarify this point, a generalization of the arguments presented in the previous section is useful.
Instead of the single observer $\bf u'$ introduced in section II along with $\bf u$, we now consider the following {\em couple} of inertial observers:
\ba
\label{u1}
u_{(1)}^\alpha&=&c\gamma_{(1)}(\delta^\alpha_0+\beta_{(1)}^j\delta^\alpha_j)\\
\label{u2}
u_{(2)}^\alpha&=&c\gamma_{(2)}(\delta^\alpha_0+\beta_{(2)}^j\delta^\alpha_j),
\ea
and focus on their relative motions.
Note that all the $\beta^j_{(n)}$ and $\gamma_{(n)}$ factors appearing in eqs.(\ref{u1}) and (\ref{u2}) are determined with respect to the same static observer $\bf u$ introduced above; their covariant form is provided by eqs.(\ref{svalpha}) and (\ref{gamma}), respectively, with the substitutions $\{u'^\alpha\rightarrow u_{(n)}^\alpha,\:\:n=1,2\}$ made.

The covariant S-velocity vector of ${\bf u}_{(2)}$ relative to ${\bf u}_{(1)}$ reads
$$
\label{s12}
\sv_{(2,1)}^\alpha=\frac{u_{(2)}^\alpha}{\gamma_{(2,1)}}-u_{(1)}^\alpha;
$$
specularly, the covariant S-velocity vector of  ${\bf u}_{(1)}$ relative to ${\bf u}_{(2)}$  is
$$
\label{s21}
\sv_{(1,2)}^\alpha=\frac{u_{(1)}^\alpha}{\gamma_{(1,2)}}-u_{(2)}^\alpha.
$$
In these formulae, the relative $\gamma$ factors $\gamma_{(1,2)}$ and $\gamma_{(2,1)}$ are given by
\be
\label{gm}
\gamma_{(1,2)}=-\frac{u_{(1)\alpha} u_{(2)}^{\alpha}}{c^2}=\gamma_{(2,1)};
\ee
in the specific case of eqs.(\ref{u1}) and (\ref{u2}), we find
$$
\label{gmb}
\gamma_{(1,2)}=\gamma_{(1)}\gamma_{(2)}
(1-\delta_{jk}\beta^j_{(1)}\beta^k_{(2)})=\gamma_{(2,1)}.
$$
Obviously, we have $\sv_{(1,2)}^\alpha\neq\sv_{(2,1)}^\alpha$, but reciprocity
implies that their moduli are equal: $\sv_{(1,2)}=\sv_{(2,1)}\equiv\sv$; specifically, we find
\be
\label{vort12}
\sv=c\sqrt{
1-\frac{(1-\beta^2_{(1)})(1-\beta^2_{(2)})}
{(1-\delta_{jk}\beta^j_{(1)}\beta^k_{(2)})^2}}=c\sqrt{1-\frac{c^4}{(u_{(1)\alpha}u_{(2)}^\alpha)^2}}\;.
\ee
Evidently, if $\beta^j_{(1)}=\beta^j_{(2)}\,\forall j$, the relative S-velocity (\ref{vort12}) is zero.
We can introduce the following
\begin{defin}
\label{def4}
Two observers are termed {\em space comoving} when their relative
S-velocity vanishes.
\end{defin}
Eq.(\ref{vort12}) shows that the relative S-velocity tends to $c$ as {\em either} of the two
observers approaches the state of a luminal particle.
Clearly, in this case, only the subluminal particle --- the $\beta<1$ one --- preserves the prerogative of representing a physical
observer; the other --- the luminal one --- loses it, and can only be regarded as an observable.
The space velocity of the latter will obviously equal $c$ regardless of the particular value of the observer's $\beta$,
{provided that} $\beta<1$.

The relative T-velocity vectors between ${\bf u}_{(1)}$ and ${\bf u}_{(2)}$ read:
\ba
\tv_{(2,1)}^\alpha&=&\frac {iu_{(1)}^\alpha}{\gamma_{(1)}\gamma_{(2)}(1-\delta_{jk}\beta^j_{(1)}
\beta^k_{(2)})}=\frac{iu^\alpha_{(1)}}{\gamma_{(2,1)}}\nonumber\\
\tv_{(1,2)}^\alpha&=&\frac {iu_{(2)}^\alpha}{\gamma_{(1)}\gamma_{(2)}
(1-\delta_{jk}\beta^j_{(1)}\beta^k_{(2)})}=\frac {iu_{(2)}^\alpha}{\gamma_{(1,2)}}\nonumber,
\ea
where the final rhs forms make covariance evident, once eq.(\ref{gm}) is recalled.
Alike the previous case, we have $\tv_{(1,2)}^\alpha\neq\tv_{(2,1)}^\alpha$,
 but their moduli are equal: $\tv_{(1,2)}=\tv_{(2,1)}\equiv\tv$, reading
\be\label{tvelocities}
\tv=\frac{c\sqrt{(1-\beta^2_{(1)})(1-\beta^2_{(2)})}}
{1-\delta_{jk}\beta^j_{(1)}\beta^k_{(2)}}=\frac{c}{\gamma_{(1,2)}}.
\ee
With the relative T-velocity (\ref{tvelocities}) thus introduced, we can now issue a parallel to definition \ref{def4} in the following
\begin{defin}
\label{def5}
Two observers are termed {\em time comoving} when their
relative T-ve\-loc\-i\-ty vanishes.
\end{defin}
In the traditional terminology, space comoving obververs are called ``comoving" {\em tout court}.
Yet, proposition \ref{pr2} -- the generalized Principle -- implies that two observers with a null relative S-velocity do {\em not} comove in time (contrary to what the usual spacetime diagrams might induce to think).
Indeed, space comoving observers are {\em maximally} non comoving in time: their relative T-velocity is $\tv=c$, as eq.(\ref{tvelocities}) shows.
In this case, each of them therefore {observes} the other to move along one's own time axis at the maximum allowed speed, i.e. $c$.

 From eqs.(\ref{vort12}) and (\ref{tvelocities}) the fundamental result stated in proposition \ref{pr1} is obviously recovered: the
world-velocity of any of the two observers as measured by the other is
$w=\sqrt{\sv^2+\tv^2}=c$, {identically}.
Since time comotion would require the relative S-velocity of the two {\em observers} to equal
$c$, it represents a physically unrealizable circumstance.
Hence, definition \ref{def5} describes just a virtual limit case: two physical observers
can only {\em approach} time comotion asymptotically, when either of them is in the
ultrarelativistic regime.
Nevertheless, this limit case is useful in that it leads to the observation that a lightlike particle would appear as time-comoving, and hence
should have no T-velocity at all, with respect to {\em any} observer --- in agreement with the results of the previous section.
Thus, from proposition \ref{pr2} it follows that any observer measuring the S-velocity of a
lightlike particle will get $c$ as the result: the experimentally verified absolute character of the space velocity of light in vacuo emerges naturally as a simple consequence of the absolute character of the world-velocity.

These observations can be collected in the following
\begin{prop}
\label{pr3}
With respect to any observer, a lightlike particle is characterized
by having $\sv=c$ and $\tv=0$, identically.
Hence, the world-velocity of a lightlike particle coincides with its S-velocity.
\end{prop}

The physical meaning of the expression ``a {null} T-velocity" is worth further remarks.
A lightlike particle, with its null T-velocity, is actually {\em observed} to behave as if it
were confined to the three space dimensions, with the time dimension made unaccessible to it.
While a lightlike particle cannot represent an observer, since it carries no clock whatever
(``no proper time for photons"), a particle characterized by a nonzero T-velocity has a definite link with the time dimension, and this holds true for any {\em non}luminal particle.
It is a well established fact that a subluminal particle cannot
be accelerated to a S-velocity equal to the velocity of light;
from the generalized Principle --- proposition \ref{pr2} --- and the absoluteness of the world-velocity,
one can equivalently say that the T-velocity of the particle cannot be forced to vanish.
As we are going to see in the following section, this circumstance would correspond to a  change of the
particle identity, since the vanishing of its T-velocity would imply a vanishing
of its mass (a scalar invariant quantity).

It might be observed that such a behaviour points to an asymmetry between the time and the space dimensions, since there seem to be neither conceptual nor
formal difficulties in setting {\em space} velocities equal to zero with respect to
a given observer, instead.
Yet, taking quantum mechanics into account, we observe that indetermination prevents the S-velocity of a particle from being exactly zero;
equivalently, we cannot observe a zero entropy--zero absolute temperature state. Realistically,
the statement ``space-velocity equal to zero" should be properly applied only with reference to an average property of a statistical ensemble of particles.
With allowance made to these non-classical considerations, and extending the concept of the spacetime continuum to the quantum regime,
we therefore see that full symmetry between the time and space dimensions is recovered again.

Since the thread connecting a nonluminal particle to the time dimension
can be streched on endlessly (with diverging $\gamma$) but cannot be cut,
we can state the following
\begin{prop}
\label{pr4}
The T-velocity of a particle cannot be forced to vanish if it is
initially nonzero, nor it can become nonzero if it is initially null.
\end{prop}
The question might then arise of why a light signal propagating through an optical medium
``slows down" to a space velocity $\sv=c/n<c$, where $n$ is the refractive index of the medium:
indeed, this would appear to violate the second statement in proposition \ref{pr4}.
The point is that the propagation of light in an optically refractive medium with $n\neq 1$
consists of a succession of {absorption} and {re-emission} processes by the atoms of the medium itself;
at each intermediate step, the velocity of the photon is {\em always} $c$, but the photon itself is {\em not} conserved along the whole path:
 the propagation of  light
throught the medium is the result of the propagation of many {\em different} photons, with an overall delay
which {mimics} that of a single light signal with a velocity $c/n$ instead of $c$.

\section{Time-velocity and mass}

The considerations made in the previous section point to the existence of a connection between the time-velocity of a particle and its mass.
Mass is a scalar invariant quantity, and therefore represents an absolute property of a particle;
as a consequence, its value cannot be altered by
any physical process preserving the particle identity.
The impossibility of
reducing the T-velocity of a nonluminal particle to zero (proposition \ref{pr4}) is related to the impossibility of reducing the
particle mass to zero, a fact which holds true for any observer.
Specularly, the identical vanishing of the
T-velocity for a lightlike particle appears to be naturally associated with its being massless.

If we consider a free massive particle $\bf u'$, and its time- and space-velocities expressed in terms of its linear momentum
${\bf p}'=m{\bf u}'$, relation (\ref{svalpha}) is observed to be mass {\em independent}:
\be
\label{svalphap}
\sv(u',u)^\alpha=-\frac{c^2P(u)^\alpha{}_\beta u'{}^{\beta}}{(u_\sigma u'{}^{\sigma})}=-\frac{c^2P(u)^\alpha{}_\beta p'{}^{\beta}}{(u_\sigma p'{}^{\sigma})}=\sv(p',u)^\alpha.
\ee
Being valid irrespective of mass, the above relation can tentatively be applied also to a photon of
four-momentum $\bf{k}$;
in this case,
with ${p'}^\alpha\equiv k^\alpha$, eq.(\ref{svalphap}) would give
$\sv_{\rm ph}=\sqrt{\sv_{\rm ph}(k,u)_\alpha\sv_{\rm ph}(k,u)^\alpha}=c$,
which actually agrees with the experimental results.
Application of the same procedure to the expression (\ref{tvalpha}) of the T-velocity vector is not so immediate, though.
Since the denominator in eq.(\ref{tvalpha})
is quadratic in the particle four-velocity, while the numerator linearly depends on it, if we rewrite eq.(\ref{tvalpha})
in terms of $\bf p'$ we obtain an expression which is explicitly mass
 {\em dependent}:
\be
\label{tvm}
\tv(p',u)^\alpha=-\frac{imc^4\Pi(u)^\alpha{}_\beta {p'}^\beta}{(u^\sigma p'_\sigma)^2}=m\tv(u',u)^\alpha,
\ee
with modulus
\be\label{modv}
\tv=-\frac{mc^3}{u^\sigma p'_\sigma}.
\ee
The quantity appearing in the denominator of eq.(\ref{modv}) is (minus) the particle total energy, as locally measured by $\bf u$ \cite{dFC}; this
 is an always nonnull quantity, which can become arbitrarily large {only asymptotically}.
 Therefore, the time-velocity can tend to zero only as a limit ---
 in so far as its mass $m$ is not null.
 This observation lets us
better appreciate proposition \ref{pr4}: the nonnull T-velocity of a nonluminal particle
cannot be forced to vanish because its invariant mass cannot vary --- and in particular cannot become null --- without changing the very
identity of the particle itself.

Applicability of relation ({\ref{tvm}) to a lightlike particle must undergo a constraint:
we cannot limit ourserves to setting ${p'}^\alpha\equiv k^\alpha$ as done above, but we must also impose $m= 0$; only
in this case, in fact, the result $\tv\equiv 0$ is obtained.
Evidently, the opposite argument can also be made: position ${p'}^\alpha\equiv k^\alpha$
provides a null T-velocity for the luminal particle only if condition $m=0$ is imposed.
These considerations allow us to state the following
\begin{prop}
\label{pr5}
With respect to any observer, a {\em massless} particle can only move with a S-velocity equal to $c$.
\end{prop}

So far, we have been dealing with particle {\em velocities}; particle {\em momenta} have been introduced only in terms of these velocities, and the lightlike behaviour has been dealt with only as a limit case.
However, a unified treatment of luminal and nonluminal particles can be done in terms of their four-momentum, without recurse to limit procedures.
This issue will occupy the oncoming section.

\section{Momentum relations}

Let $\bf u$ be the observer, and $\bf p'$ the four-momentum of the observed particle;
we do not specify if this particle follows timelike trajectories or lightlike ones: the treatment will proceed in full generality.
According to $\bf u$, the space-momentum of the particle is given by the projection of $\bf p'$
on his/her local rest frame; employing the corresponding projector (\ref{Pop}), we get
\be
\label{ps}
\ps(p',u)^\alpha=P(u)^\alpha{}_\beta p'^\beta=p'^\alpha+\frac{u_\beta p'^\beta}{c^2}\,u^\alpha.
\ee
Since
\be
\label{lE}
{\cal E}(p',u)=-u_\alpha p'^\alpha
\ee
is the energy of the particle as locally measured by $\bf u$, eq.(\ref{ps}) can be rewritten as
\be
\label{sm}
\ps(p',u)^\alpha=p'^\alpha-\frac{{\cal E}(p',u)}{c^2}\,u^\alpha.
\ee
The time-momentum of the particle is obtained by projecting $\bf p'$ on the {\em spatial} time axis of $\bf u$; this is done by employing the {\em spatial} time projector: $i\Pi(u)^\alpha{}_\beta$, cf.\,eq.(\ref{tvalpha}), where $\Pi(u)^\alpha{}_\beta$ is the time projector (\ref{Pip}).
The result is
\be
\label{tm}
\pt(p',u)^\alpha=i\Pi(u)^\alpha{}_\beta p'^\beta=-\frac{iu^\alpha u_\beta p'^\beta}{c^2}=\frac{i{\cal E}(p',u)}{c^2}u^\alpha.
\ee
Due to the intrinsic character of the projectors $P(u)^\alpha{}_\beta$ and $i\Pi(u)^\alpha{}_\beta$, eqs.(\ref{sm}) and (\ref{tm}) define {\em spacelike} vectors both; if we calculate their square moduli, from eq.(\ref{sm}) we find the relation
\be
\label{Es}
{\cal E}^2(p',u) =[\ps^2(p',u)-p'^2]\,c^2,
\ee
and from eq.(\ref{tm}) the relation
\be
\label{Et}
{\cal E}^2(p',u)=\pt^2(p',u)\,c^2 .
\ee
These two relations hold for {\em all} kinds of particles.
The specific timelike or lightlike character of these particles is assessed when a nonnull or null value, respectively, is assigned to the square modulus of their momentum, i.e., to the scalar invariant quantity $\zeta$ in the relation
\be
\label{az}
p'_\alpha p'^\alpha=-\zeta^2.
\ee
Equations (\ref{Es}) and (\ref{Et}) imply
\be
\label{zeta}
\zeta^2=\pt^2(p',u)-\ps^2(p',u)
\ee
and the {general} relation
\be
\label{genr}
{\cal E}^2(p',u)={\cal E}_{\rm kin}^2(p',u)+\zeta^2 c^2=\pt^2(p',u)\,c^2,
\ee
where we have defined
$$
{\cal E}_{\rm kin}(p',u)\equiv\ps(p',u)\,c
$$
the fraction of the measured particle energy (\ref{lE}) which is due to the {\em space} motion of the particle itself with respect to the observer $\bf u$ (in traditional terms,  the locally measured ``kinetic energy" of the particle).
Thus, we see that for all kinds of particles the time-momentum determines the total energy, while the space-momentum determines its kinetic fraction.
Note that these results do not involve the concept of ``mass", which has not been introduced yet.

If we now consider a free {\em massive} particle of mass $m$, since ${\bf p'}=m\bf{u'}$ we see that in this case $\zeta\equiv mc$, and eq.(\ref{genr}) therefore reduces to the well known relation
$$
{\cal E}^2(p',u)={\cal E}_{\rm kin}^2(p',u)+m^2c^4,
$$
while eq.(\ref{zeta}) shows that mass measures the dissymmetry between the space- and time-momenta:
\be
\label{m}
m=\sqrt{\pt^2(p',u)-\ps^2(p',u)}\,/c.
\ee
Since $m=\zeta/c$ is an invariant, eq.(\ref{m}) must provide a constant value with varying $\pt(p',u)$
and $\ps(p',u)$; in particular, if the particle is space-comoving ($\ps=0$) with the observer $\bf u$,
the particle mass fixes the minimum time-momentum a given particle can possess.
Eq.(\ref{m}) also sets a higher bound for the particle space-momentum, which must
necessarily satisfy $\ps(p',u)\le\pt(p',u)$, the equality corresponding to the lightlike particle case.
In this case, in fact, the scalar invariant quantity $\zeta$ of eq.(\ref{az}) is null, and the above mentioned dissymmetry between space- and
time-momenta is automatically removed: from eqs.(\ref{zeta}) and (\ref{genr}) we get
$$\pt(p',u)=\ps(p',u)$$
together with the well known relation
$$
{\cal E}(p',u)=\ps(p',u)\,c.
$$
The equality of space- and time- momenta for a lighlike particle is consistent with relation (\ref{m}) when applied to a massless particle.
Thus, even if the lightlike particle behaviour has not been obtained in this section as the ultrarelativistic limit of the timelike one, as done in the previous sections, it is fully consistent
with this same limit.
This reciprocally supports the consistency of the above employed limit procedure in dealing with the lightlike particle case.

As a final addendum, we can express the space- and time-momenta of a free massive particle in
terms of its space- and time-velocity vectors, in a covariant way; we obtain:
\ba
\ps(u',u)^\alpha&=&-m\frac{u_\sigma u'^\sigma}{c^2}\sv(u',u)^\alpha\nonumber\\
\pt(u',u)^\alpha&=&m\frac{(u_\sigma u'^\sigma)^2}{c^4}\tv(u',u)^\alpha\nonumber.
\ea
No corresponding relations can be provided for the case of massless particles;
note in particular that the time-momentum of a lightlike particle is not null even if its time-velocity
always is.

\section{Applications: the twin problem}

The most popular implication of the time dilation effect is the twin problem. We shall see how the correct answer stems from the very definition of T-velocity.

Consider an inertial frame $K$ spanned by cylindrical coordinates:
\be\label{1}
ds^2=-c^2dt^2+d{r}^2+{r}^2d\phi^2+dz^2
\ee
and a physical observer $\bf u$ spatially at rest in $K$, namely $u^\alpha=\delta^\alpha_0$;  $K$ is the frame of $\bf u$.
Let a twin observer $\bf u'$ leave $\bf u$ at some initial space-time position ($t=0,r=r_0,\phi=0,z=0$, say), and perform a long journey along a spatially circular orbit with
\be\label{2}
u^{'\alpha}=c\,{\text e}^\psi\left(\delta^\alpha_0+\frac{\omega}c\delta^\alpha_\phi\right)
\ee
where ${\text e}^\psi=(1-\omega^2{r}^2/c^2)^{-1/2}$ and $\omega=d\phi/dt=const.$, with $\omega r<c$; we also require $z=0$. After a $2\pi$ variation of the coordinate $\phi$, the traveller meets again her twin brother at the spatial position of departure, finding him much older than she was. The proper time lapsed on her clock is equal to
\be
\label{3}
\Delta T'_{u'}=\left(1-\frac{\omega^2 {r}^2}{c^2}\right)^{1/2}\Delta t_{u}
\ee
where $\Delta t_{u}$ can be explained following the arguments of Section II. The journey of the traveller $\bf u'$ is {\it seen} in the frame $K$, where each spacetime point carries a clock signing the time coordinate $t$ of the event. Along her journey, the traveller crosses at each point
a clock of $K$ which marks the time of crossing; this time we term $t_{u}$, having in mind that each of these
$t_{u}$'s are read on different clocks belonging to $K$. The finite quantity $\Delta t_{u}$
then is the coordinate time interval read by $\bf u'$ on the $\bf u$-clocks along her journey.  Nevertheless, from metric form (\ref{1}) and the equation of $\bf u$ we deduce that $\Delta t_u$ is also the over all proper time $\Delta T_u$ lapsed on the clock of the twin observer  $\bf u$
 who remains at the initial spatial position until he is joined by his twin sister.
Comparing this with (\ref{3}) we deduce the well known relation:
\be\label{4}
\Delta T'_{u'}=\left(1-\frac{\omega^2{r}^2}{c^2}\right)^{1/2}\Delta T_u.
\ee
This result can be obtained in a  straightforward way using the concept of time-velocity.
In the frame $K$, the twin traveller  $\bf u'$ is seen to move in the time dimension of the static observers $\bf u$'s which we know exist at each point of $K$, with a constant instantaneous T-velocity $v_T(u',u)$.
Hence, the observer $\bf u'$ is seen to cover a distance in the time dimension of $\bf u$ equal to
\be\label{5}
\Delta L_{u'}=c\Delta T'_{u'}=v_T(u',u)\Delta t_{u}.
\ee
But $v_T=c/\gamma=c{\text e}^{-\psi}$ and $\Delta t_{u}=\Delta T_u$, hence (\ref{5}) reduces to  (\ref{3}) as expected.

The above analysis, performed from the point of view of the observer $\bf u'$, is less trivial.

The observer $\bf u'$ can be considered at rest in a non inertial frame $K'$ described by the line element
\ba\label{6}
ds^2&=&-(1-\omega^2{r'}^2/c^2)c^2dt^{'2}+d{r'}^2+2\omega {r'}^2dt'd\phi'\nonumber\\
&+&d{r'}^2+{r'}^2d\phi^{'2}+dz^{'2}
\ea
where the primed coordinates span a rigidly rotating spacetime with $\omega r'<c$. $K'$ is the frame of $\bf u'$. The observer $\bf u'$ has now the form
\be\label{7}
u^{'\alpha}=c\,\delta^\alpha_0\left(1-\frac{\omega^2{r'}^2}{c^2}\right)^{-1/2},
\ee
while the twin $\bf u$ is seen to move on a spatially circular route given by
\be\label{8}
u^\alpha=c\,\left(\delta^\alpha-\frac\omega c\delta^\alpha_\phi\right).
\ee
The latter is an inertial physical observer because he is unitary (modulo $c$)  with respect to metric
(\ref{6}) and his proper time is just the coordinate time $t'$. The inertial observer will join his sister twin after a $2\pi$ cycle of the coordinate $\phi'$ relative to $\bf u'$. The proper time spent by the observer $\bf u$ for the whole journey is
\be\label{9}
\Delta T_u=-\oint_{\phi'}u_\alpha dx^{'\alpha}=\Delta t'
\ee
while the proper time spent by the twin $\bf u'$ at rest in her initial spatial position, for a coordinate time interval $\Delta t'$ and setting
$d\phi'=0$, is
\ba\label{10}
\Delta T'_{u'}=-\int u'_\alpha dx^{'\alpha}&=&\left(1-\frac{\omega^2{r'}^2}{c^2}\right)^{1/2}\Delta t'\nonumber\\
&=&\left(1-\frac{\omega^2{r'}^2}{c^2}\right)^{1/2}\Delta T_{u}
\ea
as expected.

Let us now apply the notion of time-velocity. The observer $\bf u$ is seen to move in the time dimension of $\bf u'$ with a T-velocity $v_T(u,u')=c(1-\omega^2{r'}^2/c^2)^{1/2}$. The time read on the $\bf u'$-clocks which were passed by during the journey of $\bf u$ is given by
\ba\label{11}
{\Delta t'_{u'}}=-\int_{\phi'} u'_\alpha u^\alpha dt'&=&\left(1-\frac{\omega^2{r'}^2}{c^2}\right)^{-1/2}\Delta t'\nonumber\\
&=&\left(1-\frac{\omega^2{r'}^2}{c^2}\right)^{-1}\Delta T'_{u'}
\ea
from (\ref{10}), hence the observer $\bf u$ will cover a proper distance in the time dimention of $\bf u'$  given by
\be\label{12}
\Delta L_u=c\Delta T_u=v_T(u,u'){\Delta t'_{u'}} =c\left(1-\frac{\omega^2{r'}^2}{c^2}\right)^{-1/2}\Delta T'_{u'}
\ee
as it should be.

\section{Conclusion}

Introduction of the new concept of ``world-velocity" naturally implies that such a statement as
{``light propagates in vacuum with velocity $c$ regardless
of the observer who measures it"} is but a special case of the more general rule stating that
{``every physical entity moves in spacetime with a {\em world}-velocity equal to $c$ regardless of
the observer who measures it".}
This stands as the ``generalized Principle of Relativity".
With the introduction of the {physically measurable} time- and world-velocities, full equivalence is established among the motions along all the four spacetime coordinates --- in the spirit of
 Relativity --- and
a unified scenario is evidenced, in which {\it all} the constituents of the physical world obey the generalized Principle, moving in spacetime with {spacelike} {\em world}-velocities which share the same universal modulus $c$.

\end{document}